\documentclass{moriond}
\usepackage[superscript]{cite}

\usepackage{amsmath}
\usepackage{amssymb}

\def\be{\begin{equation}}
\def\ee{\end{equation}}
\def\bea{\begin{eqnarray}}
\def\eea{\end{eqnarray}}

\def\iMPQ{1}
\def\iMainz{2}
\def\iETHZ{3}
\def\iPSI{4}
\def\iDG{5}
\def\iIFSW{6}
\def\iCOI{7}
\def\iLISBON{8}
\def\iLKB{9}
\def\iAVEIRO{10}
\def\iYALE{11}
\def\iFR{12}
\def\iTAIWAN{13}
\def\iPRINCE{14}

\newcommand{\Photo}{\includegraphics[height=35mm]{mypicture}}

\begin{document}
\vspace*{4cm}
\title{THE PROTON RADIUS PUZZLE}

\author{
J.~J.~\textsc{Krauth},$^{\iMPQ,\iMainz}$
K.~\textsc{Schuhmann},$^{\iETHZ,\iPSI,\iDG}$
M.~\textsc{Abdou Ahmed},$^{\iIFSW}$
F.~D.~\textsc{Amaro},$^{\iCOI}$
P.~\textsc{Amaro},$^{\iLISBON}$
F.~\textsc{Biraben},$^{\iLKB}$
J.~M.~R.~\textsc{Cardoso},$^{\iCOI}$
M.~L.~\textsc{Carvalho},$^{\iLISBON}$
D.~S.~\textsc{Covita},$^{\iAVEIRO}$
A.~\textsc{Dax},$^{\iYALE,\iPSI}$
S.~\textsc{Dhawan},$^{\iYALE}$
M.~\textsc{Diepold},$^{\iMPQ}$
L.~M.~P.~\textsc{Fernandes},$^{\iCOI}$
B.~\textsc{Franke},$^{\iMPQ}$
S.~\textsc{Galtier},$^{\iLKB}$
A.~\textsc{Giesen},$^{\iIFSW,\iDG}$
A.~L.~\textsc{Gouvea},$^{\iCOI}$
J.~\textsc{G{\"o}tzfried},$^{\iMPQ}$
T.~\textsc{Graf},$^{\iIFSW}$
M.~\textsc{Guerra},$^{\iLISBON}$
J.~\textsc{Haack},$^{\iMainz}$
T.~W.~\textsc{H{\"a}nsch},$^{\iMPQ}$
M.~\textsc{Hildebrandt},$^{\iPSI}$
P.~\textsc{Indelicato},$^{\iLKB}$
L.~\textsc{Julien},$^{\iLKB}$
K.~\textsc{Kirch},$^{\iETHZ,\iPSI}$
A.~\textsc{Knecht},$^{\iPSI}$
P.~\textsc{Knowles},$^{\iFR}$
F.~\textsc{Kottmann},$^{\iETHZ}$
E.-O.~\textsc{Le~Bigot},$^{\iLKB}$
Y.-W.~\textsc{Liu},$^{\iTAIWAN}$
J.~A.~M.~\textsc{Lopes},$^{\iCOI}$
L.~\textsc{Ludhova},$^{\iFR}$
J.~\textsc{Machado},$^{\iLISBON}$
C.~M.~B.~\textsc{Monteiro},$^{\iCOI}$
F.~\textsc{Mulhauser},$^{\iFR,\iMPQ}$
T.~\textsc{Nebel},$^{\iMPQ}$
F.~\textsc{Nez},$^{\iLKB}$
P.~\textsc{Rabinowitz},$^{\iPRINCE}$
E.~\textsc{Rapisarda},$^{\iPSI}$
J.~M.~F.~\textsc{dos~Santos},$^{\iCOI}$
J.~P.~\textsc{Santos},$^{\iLISBON}$
L.~A.~\textsc{Schaller},$^{\iFR}$ 
C.~\textsc{Schwob},$^{\iLKB}$
C.~I.~\textsc{Szabo},$^{\iLKB}$
D.~\textsc{Taqqu},$^{\iPSI}$
J.~F.~C.~A.~\textsc{Veloso},$^{\iAVEIRO}$
A.~\textsc{Voss},$^{\iIFSW}$
B.~\textsc{Weichelt},$^{\iIFSW}$
M.~\textsc{Willig},$^{\iMainz}$
R.~\textsc{Pohl},$^{\iMPQ,\iMainz}$
and A.~\textsc{Antognini}$^{\iMPQ,\iETHZ,\iPSI,\,}$\footnote{speaker}\\[1ex]
{\footnotesize
{$^{\iMPQ}$Max--Planck--Institut f{\"u}r Quantenoptik, Garching,
  Germany.}\quad 
{$^{\iMainz}$Johannes Gutenberg-Universit{\"a}t Mainz, QUANTUM, Institut f{\"u}r Physik \& Exzellenzcluster PRISMA, 
  Mainz, Germany.} \quad
{$^{\iETHZ}$Institute for Particle Physics, ETH Zurich, 
  Switzerland.}\quad
{$^{\iPSI}$Paul Scherrer Institute, 
  Villigen, Switzerland.}\quad
{$^{\iDG}$Dausinger \& Giesen GmbH, 
  Stuttgart, Germany.}\quad
{$^{\iIFSW}$Institut f{\"u}r Strahlwerkzeuge, Universit{\"a}t Stuttgart,
  Germany.}\quad
{$^{\iCOI}$LIBPhys, Department of Physics, University of Coimbra,
  Portugal.}\quad
{$^{\iLISBON}$ Laborat\'orio de Instrumenta\c{c}\~ao, Engenharia Biom\'edica e F\'isica da Radia\c{c}\~ao
(LIBPhys-UNL),~Departamento de F\'isica, Faculdade de Ci\^{e}ncias e Tecnologia, FCT, Universidade Nova de Lisboa, Portugal.}\quad
{$^{\iLKB}$Laboratoire Kastler Brossel, UPMC-Sorbonne Universit\'es, CNRS, }\quad 
{ENS-PSL Research University, Coll\`{e}ge de France,
Paris, France.}\quad
{$^{\iAVEIRO}$I3N, Departamento de F\'isica, Universidade de Aveiro,
Portugal.}\quad
{$^{\iYALE}$Physics Department, Yale University, New Haven, CT,
USA.}\quad
{$^{\iFR}$D\'epartement de Physique, Universit\'e de Fribourg,
  Switzerland.}\quad
{$^{\iTAIWAN}$Physics Department, National Tsing Hua University,
  Hsinchu, Taiwan.}\quad
{$^{\iPRINCE}$Department of Chemistry, Princeton University, Princeton, NJ,
 USA.}\\[2ex]
}
}

\maketitle\abstracts{High-precision measurements of the proton radius
  from laser spectroscopy of muonic hydrogen demonstrated up to six
  standard deviations smaller values than obtained from electron-proton
  scattering and hydrogen spectroscopy. The status of this
  discrepancy, which is known as the ``proton radius puzzle'' will be
  discussed in this paper, complemented with the new insights obtained
  from spectroscopy of muonic deuterium.
}

\section{The muonic hydrogen 2S-2P experiment}\label{sec:principle}
  

At the Paul Scherrer Institute (PSI), Switzerland, we performed laser
spectroscopy of the 2S-2P transition in muonic hydrogen ($\mu$p), an atom
formed by a negative muon and a proton, with 10~ppm
accuracy~\cite{Pohl:2010:Nature_mup1,Antognini:2013:Science_mup2}.
A 6$\sigma$ discrepancy corresponding to 4 transition linewidths has
been observed between the measured 2S-2P resonance frequency and its
prediction computed in the framework of bound-state quantum
electrodynamics (QED).
This prediction requires the knowledge of fundamental constants such
as the fine-structure constant, the electron mass, the muon mass etc,
but also the proton charge radius $R_p$.
Because bound-state QED is well established, this discrepancy pointed
the attention to $R_p$ which is by far the least well-known of
the needed fundamental constants~\cite{Mohr:2016:CODATA14}.


%
The principle of the experiment is as follows: a negative
low-energy (1~keV) muon beam is stopped in a low-pressure hydrogen gas
target (1~mbar, 300~K) whereby $\mu$p in a highly excited state is formed.
%
About 1\% of the formed $\mu$p atoms end-up in the 2S-state
which is metastable (with a lifetime of 1~$\mu$s at 1~mbar H$_2$ gas
pressure~\cite{Pohl:2006:MupLL2S}) and thus amenable to laser spectroscopy.
With a delay of about 1~$\mu$s after formation, the muonic atom is
illuminated by a laser pulse at a wavelength of 6.0~$\mu$m.
On resonance, the laser light induces the 2S$\rightarrow$2P transition.
The 2P state decays immediately to the ground state emitting a 2~keV
X-ray.
The number of these laser-induced X-rays 
as a function of the laser frequency is used to reveal the 2S-2P
resonance.

A fit of the resonance with a line shape model which accounts for the
energy fluctuations of the laser pulses has been used to deduce the
2S-2P transition frequency with a relative accuracy of $1\times
10^{-5}$ (corresponding to $\Gamma/30$, where $\Gamma\approx
20$~GHz is the FWHM of the transition).
From the laser frequency measured in Hz, the transition energy in meV
can be obtained using the conversion factor $h/e$ which is known with
9 significant digits~\cite{Mohr:2016:CODATA14}.
The obtained experimental value has been  compared with the
theoretical predictions~\cite{Antognini:2013:Annals}
\begin{equation}
E_\mathrm{\mu p}(2S-2P)=
206.0336(15)\mathrm{\,meV}-5.2275(10)\mathrm{\,meV/fm^2}\times R_p^2 +
0.0332(20)\mathrm{\,meV}
\label{eq:prediction}
\end{equation}
and a proton radius of $R_p=0.84087(39)$~fm has been extracted.

The first term of Eq.~\ref{eq:prediction} accounts for several 
bound-state QED contributions (radiative, recoils, binding and
relativistic corrections), the second takes into account the shift of
the energy levels caused by the finite size of the proton, and the
third -- called the two-photon exchange contribution (TPE) -- is
related with the proton polarizability.
The finite-size effect arises from the reduced Coulomb attraction when
the orbiting particle is inside the extended proton.
It scales as $R_p^2$ and depends linearly on the overlap between the
orbiting particle wave function and the nucleus which is proportional
to $m_r^3$, where $m_r$ is the reduced mass of the bound system.

Because the muon mass is 200 times larger than the electron mass, the
finite-size contribution in muonic atoms is enhanced by about $200^3$,
enabling a precise determination of $R_p$ from laser spectroscopy of
muonic hydrogen.

\section{The proton radius puzzle}\label{sec: Discussion}

\begin{figure}
  \centerline{\includegraphics[width=0.7\linewidth]{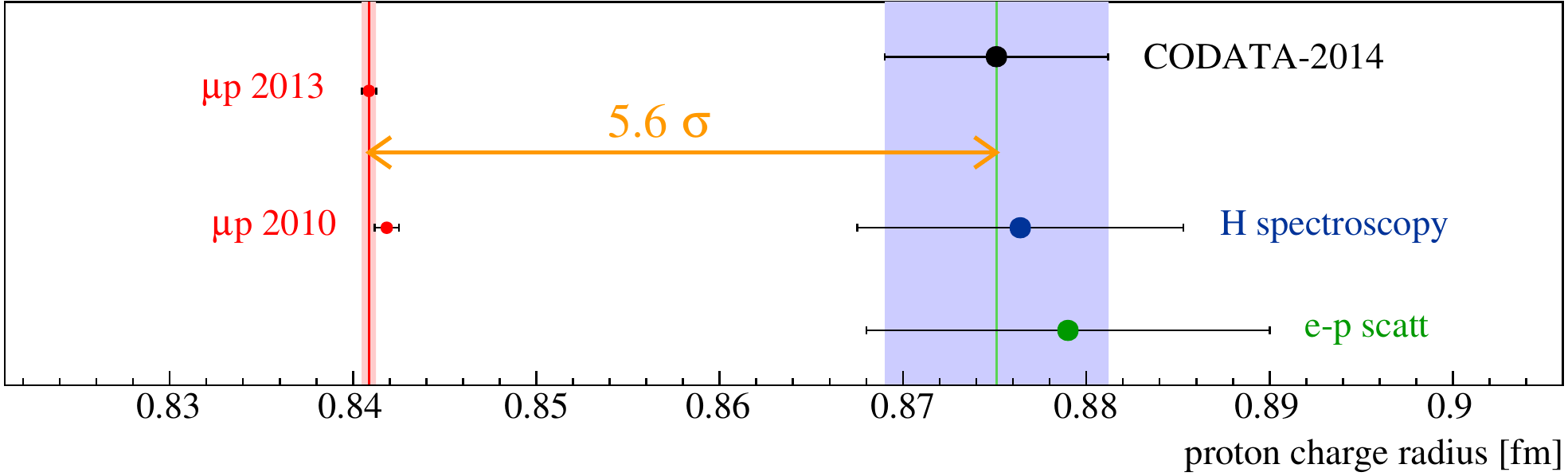}}
\caption{\label{fig:radii-propaganda} Proton charge radius from muonic
  hydrogen (red), hydrogen spectroscopy (blue) and electron-proton
  scattering (green). The CODATA value accounts for e-p scattering, H and deuterium (D)
 spectroscopy  but does not consider the muonic
  results. }
\end{figure}

There are now three methods to measure $R_p$.
The CODATA-2014 world average~\cite{Mohr:2016:CODATA14} of $R_p$ includes 
elastic scattering of electrons off protons ($e-p$)~\cite{Sick:2014:rmsRadii}
and high-precision continuous-wave laser spectroscopy of hydrogen
(H)~\cite{Biraben:2009:SpectrAtHyd,Pohl:2017:Dspec}.
%
The accuracy of $R_p$ extracted from $\mu$p surpasses the
accuracies obtained from the two other methods by an order of magnitude.
%
Yet, as visible in Fig.~\ref{fig:radii-propaganda}, a large
discrepancy exists between the muonic results and the other
determinations.
The status of this discrepancy, which is known as the ``proton radius
puzzle'',\cite{Pohl:2013:ARNPS,Carlson:2015:Puzzle,Hill:2017:PRP} will
be discussed here.
%

\subsection{Correctness of the muonic hydrogen experiment?}

Due to the $m_r^3$ scaling, the finite-size effect in $\mu$p is
strongly enhanced, while typical atomic physics systematics (e.g. the
Zeeman effect) scaling as $m_r^{-1}$ are suppressed.
Other systematic effects such as the static and dynamic Stark effects,
quantum interference~\cite{Amaro:2015:muonicQI}, pressure shift etc,
are also strongly suppressed because of the large separation between
muonic energy levels.
The hypothesis of having performed spectroscopy of
$\mu pe$ ions as suggested in~\cite{Jentschura:2011:AnnPhys2}
has been discarded
based on the measured rates and line-shapes in the muonic experiments.
In addition, many-body calculations~\cite{Karr:2012:3body,Umair:2014:molecu}
concluded that these molecular-ion states
do not exist.

So $\mu$p turns out to be very sensitive to $R_p$, but insensitive to
possible systematic effects.
The challenge in the $\mu$p experiment was to develop suitable
experimental techniques and to find the resonances.
Indeed, the rate of 6~events/h observed on resonance made the search
and thereafter the scanning of the 2S-2P resonances time consuming.
Eventually, the statistical uncertainty limited the total experimental
accuracy making the muonic results less prone to systematic results.
In this context, note that the uncertainties of the theoretical
prediction and experiment limits in almost equal parts the extraction
of $R_p$ (see also Fig.~\ref{fig:contributions}).

\subsection{Correctness of the hydrogen spectroscopy?}

The description of the energy levels in atomic H requires the
knowledge of $R_\infty$, $R_p$ and other fundamental constants.
In principle to deduce both $R_\infty$ and $R_p$ only two measurements
in H are sufficient as these other fundamental constants can be
deduced with sufficient accuracy from independent 
experiments~\cite{Mohr:2016:CODATA14}.
As ``first'' transition usually the 1S-2S transition is used, being by
far the most accurate one (relative accuracy of $4\times10^{-15}$)
and having the
largest sensitivity to $R_p$~\cite{Parthey:2011:PRL_H1S2S}. 
As ``second'' transitions, usually the 2S-$n\ell$,
with $n\ell =$ 4S, 8S, 8D, 12D, etc.\ are considered~\cite{Beauvoir:2000:Hydeurydls}.
These latter transitions have been measured with relative accuracies in the
$10^{-11}$ region and are limiting the $R_p$ extraction~\cite{Pohl:2017:Dspec}.

It turns out, that the value extracted by pairing the 1S-2S and the
2S-8D transitions is showing a $3\sigma$ deviation from $\mu$p while
all the others differ only by $\lesssim$2$\sigma$~\cite{Pohl:2013:ARNPS,Beyer:2015:Hspec}.
A $4\sigma$ discrepancy between $R_p$ from $\mu$p and H spectroscopy
alone emerges only after {\em averaging}
all measurements
in H~\cite{Mohr:2016:CODATA14}.
Therefore, a small systematic effect in these ``second''
transition measurements could be sufficient to explain the
discrepancy.
Such a systematic effect would amount to only a tiny fraction of $10^{-3}$ 
of the line width.
New measurements of $R_\infty$ are urgently needed and
underway in hydrogen
atoms~\cite{Vutha:2012:H2S2P,Beyer:2013:AdP_2S4P,Peters:2013:AdP_1S3S,Galtier:2015:1S3S}
and molecules/molecular
ions~\cite{Schiller:2014:MolClock,Dickenson:2013:H2vib,Biesheuvel:2015:HDplus,Karr:2016:HmolIon},
in
He$^+$~\cite{Herrmann:2009:He1S2S,Kandula:2011:XUV_He,Rooij:2011:HeSpectroscopy,CancioPastor:2012:3He-4He,Shiner:1995:3He,Pachucki:2012:3He
  } and in positronium~\cite{Cooke:positronium:2015}.

\subsection{Correctness of the muonic hydrogen theory?}

\begin{figure}
\centerline{\includegraphics[width=0.7\linewidth]{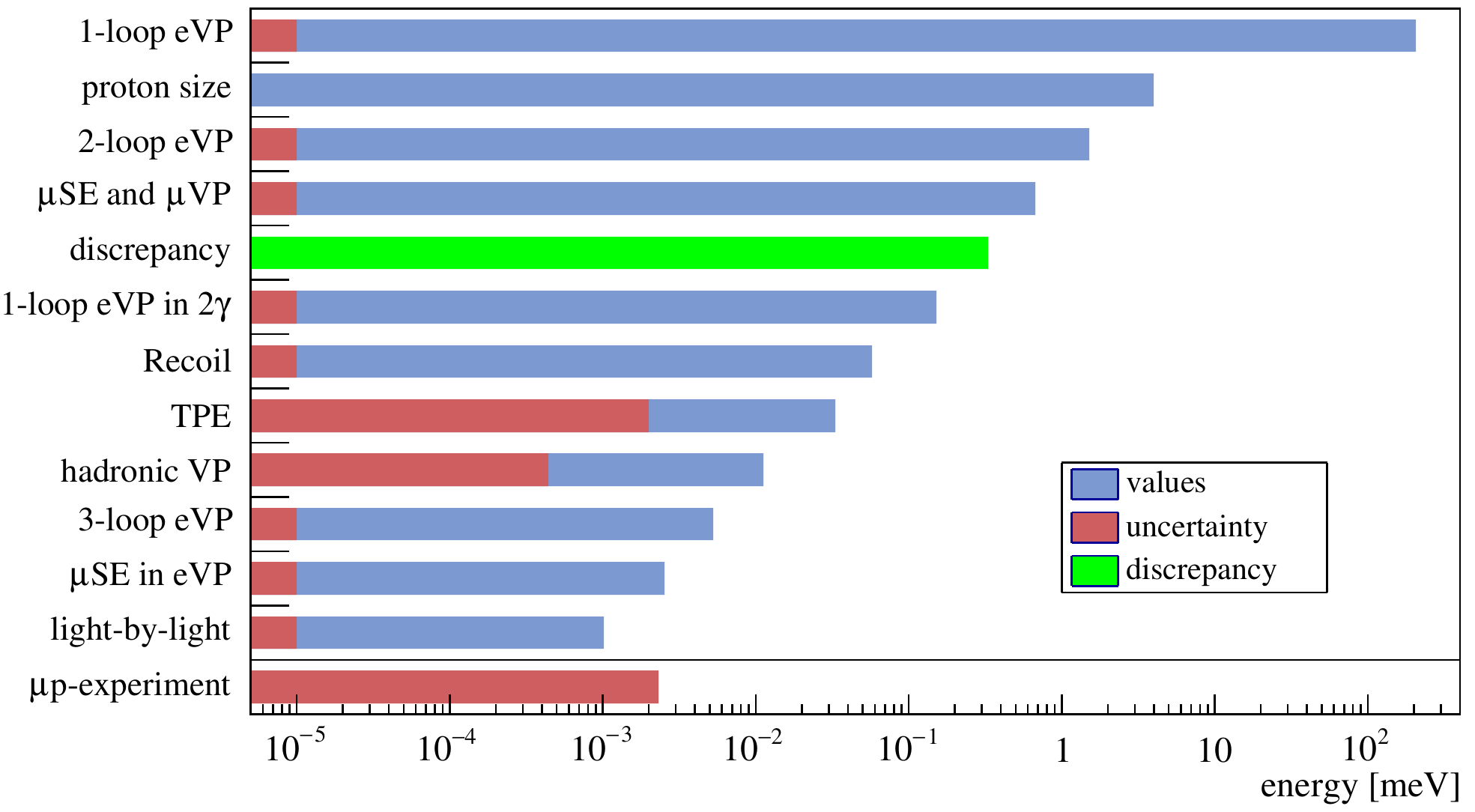}}
\caption{\label{fig:contributions} Most important contributions (blue)
  and their uncertainty (red) to the muonic hydrogen 2S-2P energy
  levels (absolute values).  Also the discrepancy (green) and the
  experimental accuracy are shown. }
\end{figure}

To extract $R_p$ from the measurement in $\mu$p we used
the prediction of Eq.~\ref{eq:prediction}\cite{Antognini:2013:Annals}.
The most important contributions underlying this equation \cite{Pachucki:1996and1999:mup,Indelicato:2012:Non_pert,Borie:2012:LS_revisited_AoP,Karshenboim:2015:mup} are shown in
Fig.~\ref{fig:contributions} together with their uncertainties.
The largest contribution is given by the one-loop electron vacuum
polarization, the second largest by the finite-size contribution.
All other contributions are thus smaller than the effect we aim to
measure.
The third largest contribution is given by the two-loop electron
vacuum polarization followed by the one-loop muon self-energy and
muon vacuum polarization.
As the discrepancy itself is larger than any other contribution, it
is highly improbable that the discrepancy  can be ascribed to
an erroneously computed or missing higher-order contribution in
$\mu$p.

Even though the TPE contribution is smaller than the discrepancy (see
Fig.~\ref{fig:contributions}), it has attracted a large interest
because it can not be simply computed using proton form factors.
It was reckoned that the uncertainty related with this hadronic
contribution and its difficult modeling could bear the solution to the
proton radius puzzle~\cite{Miller:2013:pol}.
Two approaches are used to compute this challenging contribution.
The first and most precise is an empirical approach based on
dispersion relations and measured structure functions of the
proton~\cite{Carlson:2011:PRA84_020102}, the second is based on chiral
perturbation theory ($\chi$PT)~\cite{Peset:2014:TPE}.
In the dispersive approach a subtraction
term~\cite{BirseMcGovern:2012,HillPaz:2016:Compton} is needed to cancel a divergence which
requires some modeling of the proton.
This modeling and its uncertainty was the center of a
debate~\cite{Gorchtein:2013:PRA87,Alarcon:2013:ChPT_pol_muH}.
%
Ultimately, the more recent modelings show that the uncertainty
of the subtraction term is much smaller than the discrepancy~\cite{Hill:2017:PRP, BirseMcGovern:2012,HillPaz:2016:Compton}.
The reliability of the TPE prediction has been also strongly supported
by the fact that the two approaches -- the dispersive one and  $\chi$PT based one -- give consistent results.
Thus, both the purely bound-state QED part (first term in
Eq.~\ref{eq:prediction}) and the TPE contribution (third term
in Eq.~\ref{eq:prediction}) used to extract $R_p$ from
$\mu$p are sound and have been confirmed by various groups~\cite{Pachucki:1996and1999:mup,Indelicato:2012:Non_pert,Borie:2012:LS_revisited_AoP,Karshenboim:2015:mup}.

\subsection{Correctness of electron proton scattering?}
The proton charge radius is defined as the derivative of the Sachs
electric form factor $G_E$ versus the four-momentum $Q^2$ exchanged 
\begin{equation}
R_p^2 = -6\frac{dG_E(Q^2)}{dQ^2}\Big|_{Q^2=0} \,.
\label{eq:def}
\end{equation}
This covariant definition has been applied consistently in the
description of the atomic energy levels and the electron scattering
processes~\cite{Jentschura:2011:DF}.

Because the form factor $G_E$ can be measured only down to a minimal
$Q^2$, a fit with an extrapolation to $Q^2=0$ is needed to deduce
$R_p$.
Fit functions given by truncated general series expansions such as
Taylor, Pad$\acute{\mathrm{e}}$, splines and polynomials have been 
used~\cite{Sick:2014:rmsRadii,Bernauer:2014:protonFF,Arrington:2015:escatt}: 
some authors additionally enforcing analyticity and coefficients with
perturbative scaling~\cite{Lee:2015:eScatt,HillPaz:2010:Extrapolation}, 
some others
constraining the low $Q^2$ behavior of the form factor, others
using vector meson dominance models~\cite{Beluskin:2007:dispersion, Lorenz:2015:protonFF}.
In Ref.~\cite{Sick:2014:rmsRadii} the large-$r$ behavior of the charge
distribution has been modeled by the least-bound Fock component of the
proton formed by the pion bound to a neutron ($\pi^+n$), while
in~\cite{Horbatsch:2017:mup} the higher moments of the Taylor
expansion were fixed using the higher moments of the charge
distributions predicted from $\chi$PT. 

\begin{figure}
  \centerline{\includegraphics[width=0.8\linewidth]{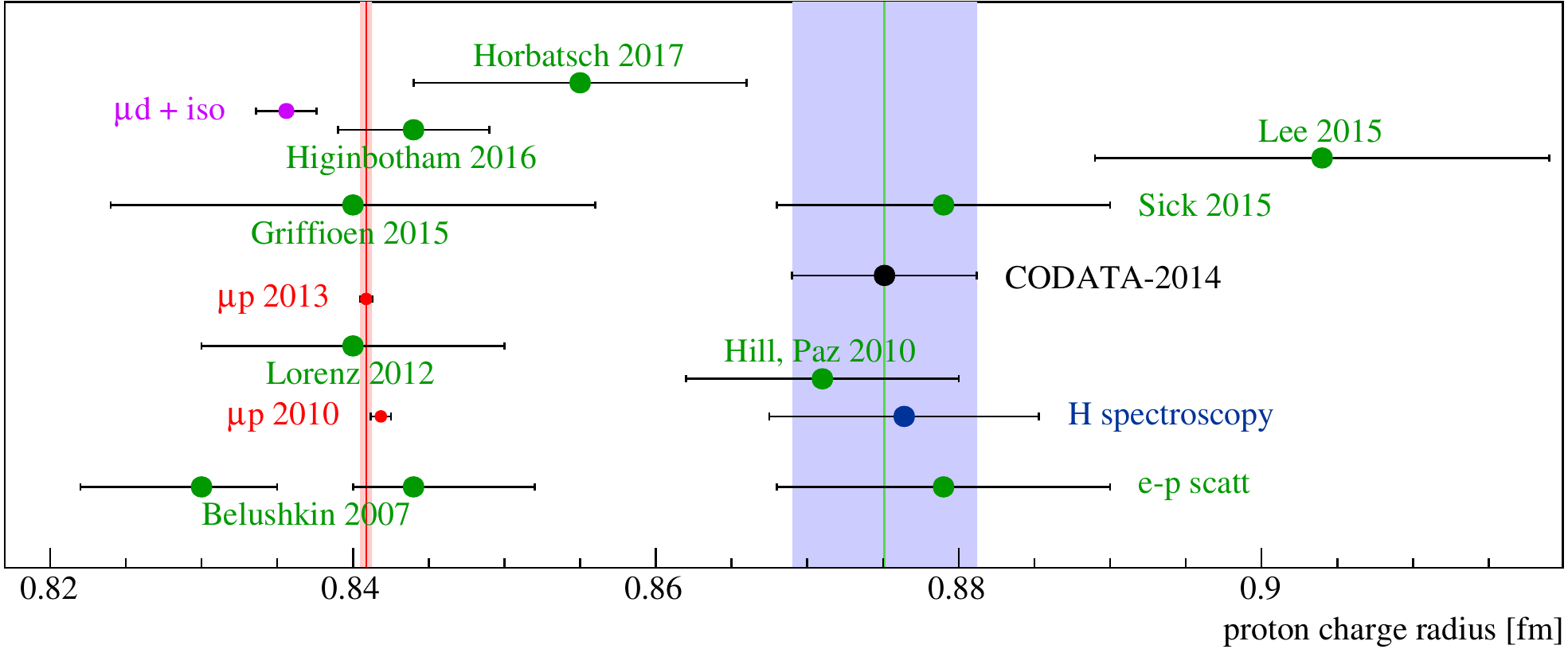}}
\caption{\label{fig:radii-many} Proton radius from muonic
  hydrogen (red), hydrogen spectroscopy (blue) and electron-proton
  scattering (green). The CODATA value does not account for the muonic
  results.}
\end{figure}

In Fig.~\ref{fig:radii-many} the most recent
$R_p$ determinations are given: some compatible with
$\mu$p~\cite{Beluskin:2007:dispersion, Lorenz:2015:protonFF,Griffioen:2016:smallPR,Higinbotham:2016:PR}, and some at
variance~\cite{Sick:2014:rmsRadii,Bernauer:2014:protonFF,Lee:2015:eScatt,Distler:2015:lowMom}.
The more traditional
analyses~\cite{Sick:2014:rmsRadii,Bernauer:2014:protonFF,Lee:2015:eScatt,Distler:2015:lowMom}
obtain $R_p$ values systematically larger than obtained by other
authors that restricted their fits to very low $Q^2$ and used
low-order power series.
Possible issues of fits restricted to very low $Q^2$ have been
demonstrated by analyzing pseudo-data generated with known
$R_p$~\cite{Kraus:2014:Polynomial,Bernauer:2016:pitfalls}.

Because data at lower $Q^2$ would facilitate the extrapolation to
$Q^2=0$, two electron-proton scattering experiments have been
initiated: one by the PRad collaboration at
JLAB~\cite{Gasparian:2011:PRad}, the other at MAMI
Mainz~\cite{Mihovilovic:2014:ISR_exp_MAMI}, 
both aiming at
$Q_\mathrm{min}^2\approx2 \times 10^{-4}$ GeV$^2$/c$^2$.
To reach the lower $Q^2$, the PRad collaboration uses of a windowless
H target and a novel non-magnetic calorimeter, the Mainz
collaboration utilized  initial state radiation.
The PRad collaboration is currently analyzing the high-quality
data collected in 2016~\cite{Gasparian:2017:PRad}.
A pilot measurement was accomplished in Mainz in 2013 demonstrating the
feasibility to extract form factors at very low $Q^2$ using initial
state radiation~\cite{Mihovilovic:2014:ISR_exp_MAMI}.
In future, a windowless hypersonic jet target is expected to reduce background
arising from the target walls and will eventually yield a competitive
$R_p$ value.

The measurement of $R_p$ using elastic {\em muon}-proton scattering at
low Q$^2$ has been proposed by the MUSE collaboration at
PSI~\cite{Gilman:2013:MUSE}.
More precisely, they plan to measure $\mu^-$-p, $\mu^+$-p, $e^+$-p, and
$e^-$-p scattering.
Despite the challenges of performing such an experiment at a secondary beam 
line with large phase-space and particle contamination,
the measurement of the cross sections of these four
channels with the same setup and beam line has two advantages.
Each individual scattering process can be used to deduce
$R_p$.
However, muon-electron universality can be best addressed by
considering the ratio between $\mu^+$-p and $e^+$-p cross sections.
Common systematic effects such as efficiencies, acceptances and
extrapolation issues are partially canceling out in the ratio.
The TPE contribution on the other hand can be measured by comparing
the scattering of $\mu^+$-p with $\mu^-$-p or  $e^+$-p with
$e^-$-p.

\subsection{Beyond standard model explanations}
Several beyond standard model (BSM) extensions have been proposed
but their majority have difficulties to resolve the discrepancy
without conflicting with low energy constraints.
Still some BSM theories able to solve the proton radius puzzle have
been formulated~\cite{Tucker-Smith:2011,Karshenboim:2014:darkForces,Carlson:2015:BSM}.
However, to avoid conflicts with other observations, these models require
fine-tuning (e.g. cancellation between axial and vector components),
and coupling preferentially to muons and protons.
Moreover they are problematic to be merged in a gauge-invariant way
into the standard model~\cite{Karshenboim:2014:darkForces,Carlson:2015:BSM}.

Other possibilities have been articulated but without clear impact on
the proton radius resolution.
Examples are
breakdown of the perturbative approach in the electron-proton
interaction at short distances~\cite{Pachucki:2014:perturb}, the
interaction with sea $\mu^+\mu^-$ and $e^+e^-$
pairs~\cite{Jentschura:2015:virtPart,Miller:2015:lepton-sea}, 
the breakdown of Lorentz
invariance~\cite{Gomes:2014:CPT}, the breakdown of
the Lamb shift expansion due to non-smooth form
factors~\cite{Hagelstein:2015:breakdown}, higher-dimensional
gravity~\cite{Dahia:2016:Xdim}, and renormalization group effects for
effective particles~\cite{Glazek:2014:calc}.

\subsection{Muonic deuterium}

Measurements in muonic deuterium ($\mu$d) have recently provided new insights.
The deuteron charge radius $R_d$ can be obtained from 
the measurements~\cite{Pohl:2016:mud}
using the prediction~\cite{Krauth:2016:mud}
\begin{equation}
E_\mathrm{\mu d}(2S-2P)= 228.7766(10)\mathrm{\,meV} - 6.1103(3)\mathrm{\,meV/fm^2}\times R_d^2 + 1.7096(200)\mathrm{\,meV}.
\label{eq:prediction_mud}
\end{equation}
Relative to $\mu$p, the finite-size effect and the TPE contribution in
$\mu$d are increased by a factor of 7 and 50, respectively.
Computation of the TPE has been greatly improved recently, using two
different techniques: ab-initio few-nucleon calculations based on
modern expressions of the nuclear
potential~\cite{Hernandez:2014:PLB736_344, Pachucki:2015:PRA91} and
the phenomenological approach based on dispersion
relations~\cite{Carlson:2014:PRA89_022504}.
Nevertheless, given its size and hadronic nature, the TPE contribution is
still the contribution having by far the largest uncertainty.

\begin{figure}
\centerline{\includegraphics[width=0.7\linewidth]{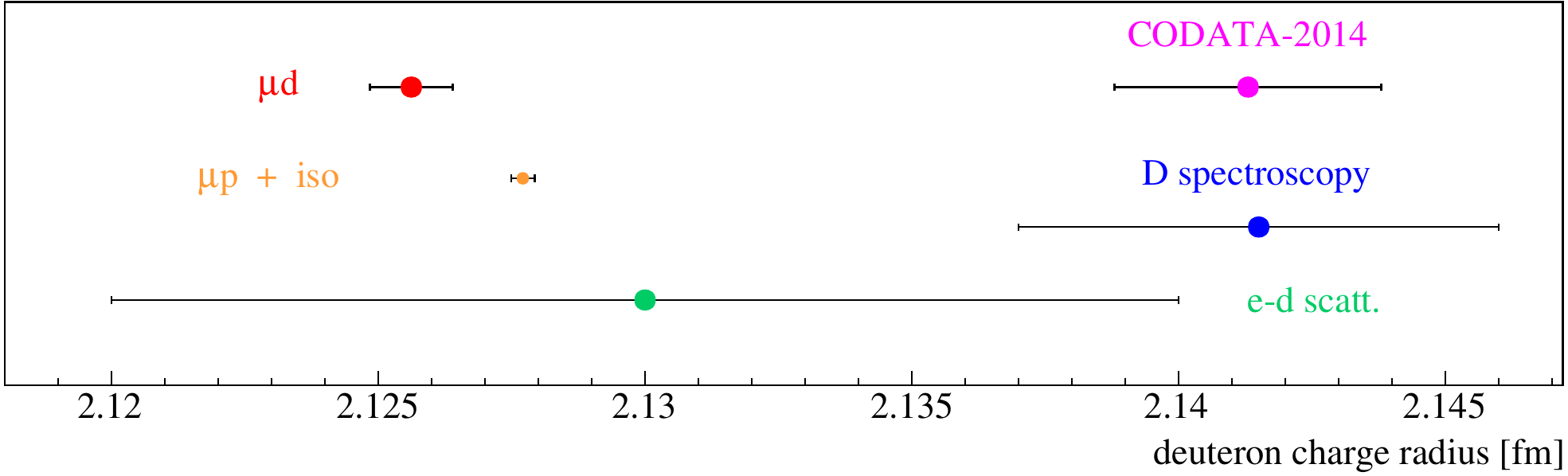}}
\caption{\label{fig:radii-deuteron} Deuteron charge radii as obtained
  from $\mu$d spectroscopy (red),  by combining $\mu$p
  spectroscopy and the H-D iso-shift measurement (brown), from
  electron scattering (green) and only D spectroscopy (blue). The
  CODATA value does not account for the muonic results but considers
  both proton and deuteron data. From~\cite{Pohl:2016:mud}.}
\end{figure}

The $R_d$ value extracted from $\mu$d spectroscopy is given in red in
Fig.~\ref{fig:radii-deuteron}.
Its error bar is dominated by the uncertainty of the TPE prediction,
while the purely QED and experimental uncertainties are negligible.
This value is in fair agreement with the value given in brown
extracted by pairing $R_p$ from $\mu$p with the difference
$R^2_\mathrm{d}-R^2_\mathrm{p}$ obtained from the measurement of the
isotopic shift of the 1S-2S transition in H and
D~\cite{Parthey:2010:PRL_IsoShift,Antognini:2013:Science_mup2}:
\begin{equation}
\begin{tabular}{cll}
  
$\left.
  \begin{array}{llrcl}
   \mbox{H-D isotopic-shift:} &R^2_\mathrm{d}-{R^2_\mathrm{p}} &  = &3.82007(65)\;\mathrm{fm}^2\\[0mm]
 \mu p:           & R_\mathrm{p}& = &0.84087(39)\;\mathrm{fm}
 \end{array}
 \hspace{-2mm}\right \}
 $
 & 
 \hspace{-0mm}$\Rightarrow $\hspace{-0mm} & ${
   R_\mathrm{d}=2.12771(22)\; \mathrm{fm\; .}} $
\label{eq:iso}
\end{tabular}
\vspace{-1ex}
\end{equation}
This agreement demonstrates the consistency of the muonic results, in
particular with the 1S-2S isotopic shift, which is a reliable
quantity.
The 2.6$\sigma$ difference could be attributed to an incomplete
treatment of the TPE contribution, or to BSM physics.
In fact, a hypothetical BSM force carrier explaining the proton radius
puzzle that does not couple to neutrons~\cite{Tucker-Smith:2011} could
produce a shift of the $\mu$d 2S-2P splitting by the observed 0.4~meV
(obtained by scaling the 0.3~meV discrepancy in $\mu$p by reduced mass
effects), explaining both muonic radii.

The $R_d$ from $\mu$d deviates by $6\sigma$ from the $R_d$ CODATA value.
The latter is however tightly linked to the {\em proton} radius
by the H-D isotopic-shift.
The blue point in Fig.~\ref{fig:radii-deuteron} is obtained from D
spectroscopy alone~\cite{Pohl:2017:Dspec}, and displays a 
$3.5\sigma$ discrepancy to the $\mu$d value.
Therefore we are facing a double discrepancy: one in the proton, the
other in the deuteron.
Unfortunately, the value from electron-deuteron scattering,
though known with a relative
accuracy of 0.5\%, is not sufficiently accurate to distinguish
between $R_d$ from D and $\mu$d.

A common solution of the atomic part of these two discrepancies could
be obtained either by changing $R_\infty$ by $6\sigma$ or the bound-state QED
theory in H/D by $50\sigma$, or by BSM physics with preferential
coupling to protons and muons.

Note that the reasoning leading to Eq.~\ref{eq:iso} can be inverted
yielding a {\em proton} radius from $R_d$ value from $\mu$d, see
Fig.~\ref{fig:radii-many}, purple point.  This confirms the small
proton radius.

\section{Conclusions}
The proton radius puzzle which to date is still unsolved has motivated
refinements of bound-state QED calculations, of theories describing
the low-energy structure of the proton~\cite{Hagelstein:2016:Review}
and deuteron, and analysis of electron scattering data.
The bound-state QED predictions agree between several authors, while
the $R_p$ extraction from electron scattering remains a controversial
subject.
Advances in the understanding of the proton structure through the TPE,
and the extrapolation to $Q^2=0$, represent an interesting platform to
test the description of analogous processes as neutrino scattering,
fundamental for the long baseline neutrino
program~\cite{Hill:2017:PRP}.
Interestingly, this advances could affect also the description of the
photon parton distribution of the proton, whose knowledge has
become important for a range of physics studies at the Large Hadron
Collider~\cite{Manohar:2016:PhysRevLett.117.242002}.
Rapid progress has been observed and is expected in the near future in
lattice QCD which offers an alternative route to the proton form
factors~\cite{Green:2014:latticeQCD,Sufian:2017:latticeQCD}.  Also
novel methods to access the radii on lattice avoiding the form factors
and its extrapolation to $Q^2=0$ are very
promising~\cite{Alexandrou:2017:latticeQCD}.
The proton radius puzzle has prompted as well several BSM extensions
studies.
Only few predictions are able to evade the various existing
constraints at the price of introducing targeted coupling, fine tuning
etc.
BSM extension could become much more natural 
if the
$R_p$ values from the atomic systems would agree and be smaller than
the value from electron scattering.
Recently, $\mu$d spectroscopy has confirmed the small value of $R_p$
as extracted from $\mu$p.
New insight from muonic helium spectroscopy, whose data analysis is
being concluded promises further insight~\cite{Antognini:2011:Conf:PSAS2010}.
Still new experimental effort in both the atomic and electron
scattering sectors which are coming online are needed to shed light
onto this puzzle.
This includes also the measurement of the ground state hyperfine
splitting in $\mu$p which is our next experimental step.

\section*{Acknowledgments}
We thank the support of the Swiss National Foundation, Projects
200021L\_138175 and 200021\_165854, of the European Research Council
ERC CoG. \#725039 and StG. \#279765, of the Deutsche
Forschungsgemeinschaft DFG\_GR\_3172/9-1, of the program PAI Germaine
de Sta\"el no. 07819NH du minist\`ere des affaires \'etrang\`eres
France, the Ecole Normale Sup\'erieure (ENS), UPMC, CNRS, and the
Funda\c{c}\~{a}o para a Ci\^{e}ncia e a Tecnologia (FCT, Portugal) and
Fundo Europeu De Desenvolvimento Regional (project
PTDC/FIS/102110/2008 and grant SFRH/BPD/46611/2008).
\section*{References}


\begin{thebibliography}{10}
\newcommand{\enquote}[1]{`#1'}
\expandafter\ifx\csname urlstyle\endcsname\relax
  \providecommand{\doi}[1]{doi:\discretionary{}{}{}#1}\else
  \providecommand{\doi}{doi:\discretionary{}{}{}\begingroup
  \urlstyle{rm}\Url}\fi

\bibitem{Pohl:2010:Nature_mup1}
R.~Pohl, A.~Antognini, F.~Nez, et~al.
\newblock \emph{Nature} \textbf{466}, 213 (2010).

\bibitem{Antognini:2013:Science_mup2}
A.~Antognini, F.~Nez, K.~Schuhmann, et~al.
\newblock \emph{Science} \textbf{339}, 417 (2013).

\bibitem{Mohr:2016:CODATA14}
P.~J. Mohr, D.~B. Newell, and B.~N. Taylor.
\newblock \emph{Rev.~Mod.~Phys.} \textbf{88}, 035009 (2016).

\bibitem{Pohl:2006:MupLL2S}
R.~Pohl, H.~Daniel, F.~J. Hartmann, et~al.
\newblock \emph{Phys.~Rev.~Lett.} \textbf{97}, 193402 (2006).

\bibitem{Antognini:2013:Annals}
A.~Antognini, F.~Kottmann, F.~Biraben, et~al.
\newblock \emph{Ann.~Phys.} \textbf{331}, 127 (2013).

\bibitem{Sick:2014:rmsRadii}
I.~Sick and D.~Trautmann.
\newblock \emph{Phys. Rev. C} \textbf{89}, 012201 (2014).

\bibitem{Biraben:2009:SpectrAtHyd}
F.~Biraben.
\newblock \emph{Eur. Phys. J. Special Topics} \textbf{172}, 109 (2009).

\bibitem{Pohl:2017:Dspec}
R.~Pohl, F.~Nez, T.~Udem, et~al.
\newblock \emph{Metrologia} \textbf{54}, L1 (2017).

\bibitem{Pohl:2013:ARNPS}
R.~Pohl, R.~Gilman, G.~A. Miller, et~al.
\newblock \emph{Ann.~Rev.~Nucl.~Part.~Sci.} \textbf{63}, 175 (2013).

\bibitem{Carlson:2015:Puzzle}
C.~E. Carlson.
\newblock \emph{Prog.~Part.~Nucl.~Phys.} \textbf{82}, 59  (2015).

\bibitem{Hill:2017:PRP}
R.~J. Hill.
\newblock \emph{arXiv:1702.01189} (2017).

\bibitem{Amaro:2015:muonicQI}
P.~Amaro, B.~Franke, J.~J. Krauth, et~al.
\newblock \emph{Phys. Rev. A} \textbf{92}, 022514 (2015).

\bibitem{Jentschura:2011:AnnPhys2}
U.~D. Jentschura.
\newblock \emph{Ann.~Phys.} \textbf{326}, 516 (2011).

\bibitem{Karr:2012:3body}
J.-P. Karr and L.~Hilico.
\newblock \emph{Phys.~Rev.~Lett.} \textbf{109}, 103401 (2012).

\bibitem{Umair:2014:molecu}
M.~Umair and S.~Jonsell.
\newblock \emph{J.~Phys.~B} \textbf{47}, 175003 (2014).

\bibitem{Parthey:2011:PRL_H1S2S}
C.~G. Parthey, A.~Matveev, J.~Alnis, et~al.
\newblock \emph{Phys.~Rev.~Lett.} \textbf{{107}}, {203001} ({2011}).

\bibitem{Beauvoir:2000:Hydeurydls}
B.~de~Beauvoir, C.~Schwob, O.~Acef, et~al.
\newblock \emph{Eur.~Phys.~J.~D} \textbf{12}, 61 (2000).

\bibitem{Beyer:2015:Hspec}
A.~Beyer, L.~Maisenbacher, K.~Khabarova, et~al.
\newblock \emph{Physica Scripta} \textbf{2015}, 014030 (2015).

\bibitem{Vutha:2012:H2S2P}
A.~C. Vutha, N.~Bezginov, I.~Ferchichi, et~al.
\newblock \emph{Bull. Am. Phys. Soc.} \textbf{57(5)}, Q1.138 (2012).

\bibitem{Beyer:2013:AdP_2S4P}
A.~Beyer, J.~Alnis, K.~Khabarova, et~al.
\newblock \emph{Ann. d. Phys. (Berlin)} \textbf{525}, 671 (2013).

\bibitem{Peters:2013:AdP_1S3S}
E.~Peters, D.~C. Yost, A.~Matveev, et~al.
\newblock \emph{Ann. d. Phys. (Berlin)} \textbf{525}, L29 (2013).

\bibitem{Galtier:2015:1S3S}
S.~Galtier, H.~Fleurbaey, S.~Thomas, et~al.
\newblock \emph{J.~Phys.~Chem.~Ref.~Data} \textbf{44}, 031201 (2015).

\bibitem{Schiller:2014:MolClock}
S.~Schiller, D.~Bakalov, and V.~I. Korobov.
\newblock \emph{Phys.~Rev.~Lett.} \textbf{113}, 023004 (2014).

\bibitem{Dickenson:2013:H2vib}
G.~D. Dickenson, M.~L. Niu, E.~J. Salumbides, et~al.
\newblock \emph{Phys.~Rev.~Lett.} \textbf{110}, 193601 (2013).

\bibitem{Biesheuvel:2015:HDplus}
J.~Biesheuvel, J.-P. Karr, L.~Hilico, et~al.
\newblock \emph{Nature Comm.} \textbf{7}, 10385 (2015).

\bibitem{Karr:2016:HmolIon}
J.~Karr, L.~Hilico, J.~Koelemeij, et~al.
\newblock \emph{arXiv 1605.05456 (physics.atom-ph)} (2016).

\bibitem{Herrmann:2009:He1S2S}
M.~Herrmann, M.~Haas, U.~Jentschura, et~al.
\newblock \emph{Phys.~Rev.~A} \textbf{79}, 052505 (2009).

\bibitem{Kandula:2011:XUV_He}
D.~Z. Kandula, C.~Gohle, T.~J. Pinkert, et~al.
\newblock \emph{Phys.~Rev.~A} \textbf{84}, 062512 (2011).

\bibitem{Rooij:2011:HeSpectroscopy}
R.~{van Rooij}, J.~S. Borbely, J.~Simonet, et~al.
\newblock \emph{Science} \textbf{333}, 196 (2011).

\bibitem{CancioPastor:2012:3He-4He}
P.~{Cancio Pastor}, L.~Consolino, G.~Giusfredi, et~al.
\newblock \emph{Phys.~Rev.~Lett.} \textbf{108}, 143001 (2012).

\bibitem{Shiner:1995:3He}
D.~Shiner, R.~Dixson, and V.~Vedantham.
\newblock \emph{Phys.~Rev.} \textbf{74}, 3553 (1995).

\bibitem{Pachucki:2012:3He}
K.~Pachucki, V.~A. Yerokhin, and P.~{Cancio Pastor}.
\newblock \emph{Phys.~Rev.~A} \textbf{85}, 042517 (2012).

\bibitem{Cooke:positronium:2015} D. A. Cooke, P. Crivelli, J. Alnis, et al.
\emph{ Hyperfine Interact.} \textbf{233}, 67 (2015).

\bibitem{Pachucki:1996and1999:mup}
K.~Pachucki.
\newblock \emph{Phys.~Rev.~A} \textbf{53}, 2092 (1996).
\newblock {\textit{Phys. Rev. A} \textbf{60}, 3593 (1999)}.

\bibitem{Indelicato:2012:Non_pert}
P.~Indelicato.
\newblock \emph{Phys.~Rev.~A} \textbf{87}, 022501 (2013).

\bibitem{Borie:2012:LS_revisited_AoP}
E.~Borie.
\newblock \emph{Ann.~Phys.} \textbf{327}, 733 (2012).
\newblock {and arXiv: 1103.1772-v7}.

\bibitem{Karshenboim:2015:mup}
S.~G. Karshenboim, E.~Y. Korzinin, et~al.
\newblock \emph{J.~Phys.~Chem.~Ref.~Data} \textbf{44}, 031202 (2015).

\bibitem{Miller:2013:pol}
G.~A. Miller.
\newblock \emph{Phys.~Lett.~B} \textbf{718}, 1078 (2013).

\bibitem{Carlson:2011:PRA84_020102}
C.~E. Carlson and M.~Vanderhaeghen.
\newblock \emph{Phys.~Rev.~A} \textbf{84}, 020102(R) (2011).

\bibitem{Peset:2014:TPE}
C.~Peset and A.~Pineda.
\newblock \emph{Nucl.~Phys.~B} \textbf{887}, 69  (2014).

\bibitem{BirseMcGovern:2012}
M.~C. Birse and J.~A. {McGovern}.
\newblock \emph{Eur.~Phys.~J.~A} \textbf{48}, 120 (2012).

\bibitem{HillPaz:2016:Compton}
R.~J. Hill and G.~Paz.
\newblock \emph{{arXiv:1611.09917}} (2016).

\bibitem{Gorchtein:2013:PRA87}
M.~Gorchtein, F.~J. Llanes-Estrada, et~al.
\newblock \emph{Phys. Rev. A} \textbf{87}, 052501 (2013).

\bibitem{Alarcon:2013:ChPT_pol_muH}
J.~M. Alarc{\'o}n, V.~Lensky, and V.~Pascalutsa.
\newblock \emph{Eur.~Phys.~J.~C} \textbf{74}, 2852 (2014).

\bibitem{Jentschura:2011:DF}
U.~D. Jentschura.
\newblock \emph{Eur.~Phys.~J.~D} \textbf{61}, 7 (2011).

\bibitem{Bernauer:2014:protonFF}
J.~C. Bernauer, M.~O. Distler, J.~Friedrich, et~al.
\newblock \emph{Phys. Rev. C} \textbf{90}, 015206 (2014).

\bibitem{Arrington:2015:escatt}
J.~Arrington and I.~Sick.
\newblock \emph{J.~Phys.~Chem.~Ref.~Data} \textbf{44}, 031204 (2015).

\bibitem{Lee:2015:eScatt}
G.~Lee, J.~R. Arrington, and R.~J. Hill.
\newblock \emph{Phys. Rev. D} \textbf{92}, 013013 (2015).

\bibitem{HillPaz:2010:Extrapolation}
R.~J. Hill and G.~Paz.
\newblock \emph{Phys.~Rev.~D} \textbf{82}, 113005 (2010).

\bibitem{Lorenz:2015:protonFF}
I.~T. Lorenz, U.-G. Mei\ss{}ner, H.-W. Hammer, et~al.
\newblock \emph{Phys. Rev. D} \textbf{91}, 014023 (2015).

\bibitem{Beluskin:2007:dispersion}
M. Belushkin, H.-W. Hammer, and U.-G. Meissner, \emph{Phys. Rev. C} \textbf{75}, 035202 (2007).

\bibitem{Horbatsch:2017:mup}
M.~Horbatsch, E.~A. Hessels, and A.~Pineda.
\newblock \emph{Phys. Rev. C} \textbf{95}, 035203 (2017).

\bibitem{Griffioen:2016:smallPR}
K.~Griffioen, C.~Carlson, and S.~Maddox.
\newblock \emph{Phys. Rev. C} \textbf{93}, 065207 (2016).

\bibitem{Higinbotham:2016:PR}
D.~W. Higinbotham, A.~A. Kabir, V.~Lin, et~al.
\newblock \emph{Phys. Rev. C} \textbf{93}, 055207 (2016).

\bibitem{Distler:2015:lowMom}
M.~O. Distler, T.~Walcher, and J.~C. Bernauer.
\newblock \emph{arXiv:1511.00479} (2015).

\bibitem{Kraus:2014:Polynomial}
E.~Kraus, K.~E. Mesick, A.~White, et~al.
\newblock \emph{Phys. Rev. C} \textbf{90}, 045206 (2014).

\bibitem{Bernauer:2016:pitfalls}
J.~C. Bernauer and M.~O. Distler.
\newblock \emph{arXiv:1606.02159} (2016).

\bibitem{Gasparian:2011:PRad}
A.~Gasparian.
\newblock \emph{EPJ Web Conf.} \textbf{73}, 07006 (2014).

\bibitem{Mihovilovic:2014:ISR_exp_MAMI}
M.~Mihovilovi{\v{c}}, H.~Merkel, A.~Weber, et~al.
\newblock \emph{EPJ Web Conf.} \textbf{72}, 00017 (2014).

\bibitem{Gasparian:2017:PRad}
A.~H. Gasparian.
\newblock \emph{JPS Conf. Proc.} \textbf{13}, 020052 (2017).

\bibitem{Gilman:2013:MUSE}
R.~Gilman.
\newblock \emph{{AIP}\ Conf.\ Proc.} \textbf{1563}, 167 (2013).

\bibitem{Tucker-Smith:2011}
D.~Tucker-Smith and I.~Yavin.
\newblock \emph{Phys. Rev. D} \textbf{83}, 101702 (2011).

\bibitem{Karshenboim:2014:darkForces}
S.~G. Karshenboim, D.~McKeen, and M.~Pospelov.
\newblock \emph{Phys. Rev. D} \textbf{90}, 073004 (2014).

\bibitem{Carlson:2015:BSM}
C.~E. Carlson and M.~Freid.
\newblock \emph{Phys. Rev. D} \textbf{92}, 095024 (2015).

\bibitem{Pachucki:2014:perturb}
K.~Pachucki and K.~A. Meissner.
\newblock \emph{arXiv:1405.6582} (2014).

\bibitem{Jentschura:2015:virtPart}
U.~D. Jentschura.
\newblock \emph{Phys. Rev. A} \textbf{92}, 012123 (2015).

\bibitem{Miller:2015:lepton-sea}
G.~A. Miller.
\newblock \emph{Phys. Rev. C} \textbf{91}, 055204 (2015).

\bibitem{Gomes:2014:CPT}
A.~H. Gomes, V.~A. Kosteleck\'y, and A.~J. Vargas.
\newblock \emph{Phys. Rev. D} \textbf{90}, 076009 (2014).

\bibitem{Hagelstein:2015:breakdown}
F.~Hagelstein and V.~Pascalutsa.
\newblock \emph{Phys. Rev. A} \textbf{91}, 040502 (2015).

\bibitem{Dahia:2016:Xdim}
F.~Dahia and A.~S. Lemos.
\newblock \emph{Eur.~Phys.~J.~C} \textbf{76}, 435 (2016).

\bibitem{Glazek:2014:calc}
S.~D. G\l{}azek.
\newblock \emph{Phys. Rev. D} \textbf{90}, 045020 (2014).

\bibitem{Pohl:2016:mud}
R.~Pohl, F.~Nez, L.~M.~P. Fernandes, et~al.
\newblock \emph{Science} \textbf{353}, 669 (2016).

\bibitem{Krauth:2016:mud}
J.~J. Krauth, M.~Diepold, B.~Franke, et~al.
\newblock \emph{Ann.~Phys.} \textbf{366}, 168  (2016).

\bibitem{Hernandez:2014:PLB736_344}
O.~Hernandez, C.~Ji, S.~Bacca, et~al.
\newblock \emph{Phys.~Lett.~B} \textbf{736}, 344 (2014).

\bibitem{Pachucki:2015:PRA91}
K. Pachucki and A. Wienczek.
\newblock   \emph{Phys. Rev. A} \textbf{91}, 040503(R) (2015).

\bibitem{Carlson:2014:PRA89_022504}
C.~E. Carlson, M.~Gorchtein, and M.~Vanderhaeghen.
\newblock \emph{Phys.~Rev.~A} \textbf{89}, 022504 (2014).

\bibitem{Parthey:2010:PRL_IsoShift}
C.~G. Parthey, A.~Matveev, J.~Alnis, et~al.
\newblock \emph{Phys.~Rev.~Lett.} \textbf{{104}}, {233001} ({2010}).

\bibitem{Hagelstein:2016:Review}
F. Hagelstein, R. Miskimen, V. Pascalutsa. 
\newblock \emph{Prog. Part. Nucl. Phys.} \textbf{88}, 29 (2016). 


\bibitem{Manohar:2016:PhysRevLett.117.242002}
A.~Manohar, P.~Nason, G.~P. Salam, et~al.
\newblock \emph{Phys. Rev. Lett.} \textbf{117}, 242002 (2016).

\bibitem{Green:2014:latticeQCD}
J.~R. Green, J.~W. Negele, A.~V. Pochinsky, et~al.
\newblock \emph{Phys. Rev. D} \textbf{90}, 074507 (2014).

\bibitem{Sufian:2017:latticeQCD}
R.~S. Sufian, Y.-B. Yang, J.~Liang, et~al.
\newblock \emph{arXiv:1705.05849} (2017).

\bibitem{Alexandrou:2017:latticeQCD}
C.~Alexandrou.
\newblock \emph{EPJ Web Conf.} \textbf{137}, 01004 (2017).

\bibitem{Antognini:2011:Conf:PSAS2010}
A.~Antognini, F.~Nez, F.~D. Amaro, et~al.
\newblock \emph{Can.~J.~Phys.} \textbf{89}, 47 (2010).

\end{thebibliography}

\end{document}